\documentclass[aps,superscriptaddress,showpacs,notitlepage,prd]{revtex4-1}
\usepackage{graphicx}
\usepackage{subfigure}
\usepackage{grffile}
\usepackage{natbib}
\usepackage{amssymb}
\usepackage{color}
\newcommand{\beq}{\begin{equation}}
\newcommand{\eeq}{\end{equation}}

\begin{document}

\title{Stability of the accretion of a ghost condensate onto the Schwarzschild black hole}

\author{Claudia A. Rivasplata Paz, Jose Martins Salim}
\affiliation{Instituto de Cosmologia, Relatividade, e Astrof\'{\i}sica ICRA - CBPF,
Rua Dr. Xavier Sigaud 150 Urca, Rio de Janeiro, Brasil.}

\author{Santiago Esteban Perez Bergliaffa}
\affiliation{Departamento de F\'{\i}sica Te\'orica, Instituto de F\'{\i}sica, 
Universidade do Estado do Rio de Janeiro, Rua S\~ao Francisco Xavier 524, Maracan\~a, Rio de Janeiro, Brasil.}

\begin{abstract}

We study the linear stability of a nongravitating, steady-state, spherically symmetric ghost condensate 
accreting onto a Schwarzschild black hole
using two methods. The first one is based on the conservation of the energy-momentum tensor of the perturbations (whose propagation 
is determined by the effective metric), and involves determination of the sign of the time derivative of the energy of the perturbations. 
The second method employs the 
positivity of the effective potential. Both methods yield the result that the system is stable, but the second one is less practical, since it involves lengthier calculations and requires the explicit form of the background solution for the scalar field. 

\end{abstract}

\pacs{completar}

\maketitle

\section{Introduction}

Black hole solutions of General Relativity in four dimensions have been shown to be dynamically stable at the linear level
by perturbing the corresponding equations of motion. 
In the case of Schwarszchild's solution, the stability was originally studied by Regge and Wheeler \citep{Regge1957}, while the stability of the Reissner-Nordstrom geometry was determined in \citep{Moncrief1974a, Moncrief1974b}. Kerr's solution
was shown to be stable in \citep{Whiting1989}. Due to the fact that 
black holes are often surrounded by matter, it is important to study 
also the stability of solutions in which there is matter accreting onto the black hole, as well as onto compact objects in general. 
The hypothesis of spherically symmetric accretion is frequently adopted as a first step in the building of more realistic models.
In this vein, the accretion of a barotropic fluid with this symmetry onto a star was studied in the newtonian regime by \citep{Bondi1952}, and its stability proved in several papers (see for instance \citep{Garlick1979} and \citep{Petterson1980}, and the references cited in \citep{Mandal2007}).
The equations of motion for a steady-state spherical symmetric flow onto a relativistic compact object were set and solved in \citep{Michel1972}.
The stationary, spherically symmetric, polytropic and inviscid accretion flow onto a Schwarzschild black hole was studied with techniques from dyamical systems in \citep{Mandal2007} \footnote{In all these publications, the fluid was taken as non-gravitating. For 
general-relativistic spherically symmetric steady accretion of self-gravitating perfect fluid onto compact objects see \citep{Mach2009}.}.
The perturbations of a non-self gravitating perfect fluid in potential flow and stationary, spherical accretion onto a Schwarzschild black hole were shown to be stable at the linear level by Moncrief \citep{Moncrief1980}, using the fact that 
their evolution is governed by what he called the sonic metric (and is now known to be a specific instance of the effective metric, see for instance \citep{Barcelo2005}). 

We shall study here the linear stability of the accretion of a scalar field 
with a non-canonical kinetic term onto a Schwarzschild black hole. 
These scalar field models, generically called $k$-essence, have been 
widely used to describe the accelerated expansion of the universe according to the standard cosmological model \citep{Armendariz2000, Armendariz2000b, Tsujikawa2010}, and also to give a unified model for dark matter and dark energy
\citep{Scherrer2004}.
Cosmological perturbations in the 
so-called 
$k$-inflation
have been analyzed in \citep{Garriga1999}.
In 
\citep{Akhoury2011b},
the gravitational collapse of
a Born-Infeld-like $k$-essence model was investigated. The collapse with other Lagrangians was studied using Painlev\`e-Gullstrand coordinates in
\citep{Leonard2011}.
Regarding black holes, 
their existence with a non-canonical scalar field as a matter source
was studied in \citep{Graham2014}. The steady-state accretion of a $k$-essence field onto a Reissner-Nordstrom black hole was studied in 
\citep{Babichev2008}. 
It was proved in \citep{Akhoury2008} that 
the existence of stationary configurations of a non-canonical scalar field requires that the Lagrangian be invariant with
respect to the field redefinition $\phi \rightarrow \phi + c$,  for all values of the constant $c$.
The steady-state spherically symmetric accretion of a ghost condensate in the gravitational field of a Schwarzschild black hole was studied analytically in \citep{Frolov2004}.
Numerical simulations of the accretion onto a Schwarzschild black hole
of test scalar fields 
with 
a Dirac-Born-Infeld type Lagrangian, and for the ghost condensate
of \citep{Frolov2004} 
were analyzed in \citep{Akhoury2011a}.

The main goal of this paper is to study the stability of the stationary accretion presented in \citep{Frolov2004} with the method developed in 
\citep{Moncrief1980} \footnote{This method was recently used to study the 
stability of 
stationary, relativistic Bondi-type accretion in Schwarzschild anti deSitter spacetimes \citep{Mach2013b}.}.
We shall begin in Sect.\ref{efmet} with a short review of the effective metric
in the case of a nonlinear field theory for a scalar field, including  
the method developed by Moncrief adapted to the case at hand.
In Sect.\ref{fro} the specific model of a nonlinear scalar field accreting onto a black hole presented in \citep{Frolov2004} will be reviewed, and shown to be stable. 
In Section \ref{efpot} we shall obtain the same result (though with a lot more effort from the point of view of calculations) using the more traditional method of the effective potential.
We close with a discussion in Sect.\ref{disc}.

\section{Effective metric}
\label{efmet}
The propagation of the excitations of any nonlinear field theory on a fixed 
background is governed by an effective metric that depends on the background field configuration and on the details
of the nonlinear dynamics obeyed by the field (see \citep{Barcelo2005} for a complete review). The case of a nonlinear scalar field has been studied in 
\citep{Barcelo2001}, and it was shown in \citep{Goulart2011} that the effective metric can be classified in different types according to whether
the gradient of the scalar field is timelike, null, or spacelike.
Let us briefly review how the effective metric arises in theories in which the action is given by
\begin{equation}
S[\phi]=\int
\sqrt{-g}\;
\mathcal{L}
(W)\;d^{4}x,
\label{action}
\end{equation}
where 
$
W\equiv g^{\mu\nu}
\nabla_\mu\phi \nabla_\nu\phi
$
and $g_{\mu\nu}$ is the background metric. 
The corresponding EOM 
is 
\begin{equation}%
\left[\sqrt{-g}\; (\mathcal{L}_{W}g^{\mu\nu} + 2\mathcal{L}
_{WW}
\Phi^{\mu\nu} )\;\right]\nabla_{\mu}\nabla_{\nu}\phi
=0,
\label{eom}
\end{equation}
where ${\cal L}_W\equiv \frac{\partial{\cal L}}{\partial W}$, etc, and
$
\Phi^{\mu\nu} \equiv
g^{\mu\alpha}\phi_{\alpha}g^{\nu\beta}\phi_{\beta}.
$
By perturbing the EOM with 
\begin{equation}
\phi=\phi_{0}+\varepsilon\phi_{1},
\end{equation}
where $\phi_{0}$ is the backgound field (solution of Eqn.(\ref{eom}) for a given Lagrangian), $\phi_{1}$ is the perturbation, and $\varepsilon$ a parameter for power counting, we get for the EOM of the perturbations (see details in \cite{Barcelo2001,Goulart2011})
\begin{equation}
\left[\sqrt{-g}
\right.(%
\mathcal{L}%
_{W}g^{\mu\nu}+2%
\mathcal{L}%
_{WW}\Phi^{\mu\nu}
)\left|_0\;
\phi_{1,\mu}\right],_{\nu}=0,
\label{eomperts}
\end{equation}
where all the quantities in the parentheses are evaluated 
at the background.
Defining 
$\widetilde{g}^{\mu\nu}$ by
\begin{equation}
\sqrt{-\widetilde{g}}\text{
}\widetilde{g}^{\mu\nu} =
\sqrt{-g}\left.\left(
\mathcal{L}%
_{W}g^{\mu\nu}+2%
\mathcal{L}%
_{WW}\Phi^{\mu\nu}\right)\right|_0,
\label{defeffmet}
\end{equation}
Eqn.(\ref{eomperts}) can be written as 
\begin{equation}
(\sqrt{-\widetilde{g}}\text{ }\widetilde{g}^{\mu\nu}\phi_{1,\mu}),_{\nu}=0.
\end{equation}
It follows from Eqn.(\ref{defeffmet}) that
\begin{equation}
\widetilde{g}^{\mu\nu}=\frac{M^{\mu\nu}}{\sqrt{-g}\sqrt{M}},
\end{equation}
where
$M^{\mu
\nu}\equiv%
\left.
\mathcal{L}%
_{W}g^{\mu\nu}+2%
\mathcal{L}%
_{WW}\Phi^{\mu\nu}\right|_0$.
Different aspects of the effective geometry for scalar fields with noncanonical Lagrangians, described by $\widetilde g_{\mu\nu}$, {\emph i.e.} the inverse of the tensor defined in Eqn.(\ref{defeffmet}), were studied in \citep{Babichev2006, Babichev2007, Goulart2011}.
Among them, we would like to point out here that the effective metric inherits the symmetries of the background metric. 
This can be seen as follows. If $X$ be a Killing vector of the backgroud metric, then $\pounds_Xg_{\mu\nu} = 0$. Since 
$\pounds_X\phi_0 = 0$, the effective metric satisfies $\pounds_X \widetilde{g}_{\mu\nu}=0$.

By perturbing the action given in Eqn.(\ref{action}), it follows that the action for the perturbations (corresponding to the $O(\varepsilon^2)$ term) reads
\begin{equation}
S_{2}=\int\sqrt{-\widetilde{g}}\text{ }\widetilde{g}^{\mu\nu}\phi_{1,\mu}%
\phi_{1,\nu}d^{4}x.
\end{equation}
The variation of this action w.r.t. 
$\widetilde{g}^{\mu\nu}$ yields the energy-momentum tensor of the perturbations,
given by
\begin{equation}
\widetilde{T}_{\nu}^{\mu}=\widetilde{g}^{\mu\lambda}\phi_{1,\lambda}%
\phi_{1,\nu}-\frac{1}{2}\delta_{\nu}^{\mu}\widetilde{g}^{\alpha\beta}%
\phi_{1,\alpha}\phi_{1,\beta},
\end{equation}
which satisfies 
$\widetilde{\nabla}_{\mu}\widetilde{T}^{\mu}_{\nu}=0$, 
where $\widetilde{\nabla}_{\mu}$ is the covariant derivative defined 
with the efective metric. 

As discussed in \citep{Moncrief1980}, if $X$ is a Killing vector of the effective metric it follows that
\begin{eqnarray}
\widetilde{\nabla}_{\mu}\left(  X^{\nu}\widetilde{T}_{\nu}^{\mu}\right)  =0,
\end{eqnarray}
which can be rewritten as 
\begin{equation}
\partial_\nu\left( \sqrt{-\widetilde{g}} {X}^{\mu}\widetilde{T}_{\mu
}^{\nu}\right) = 0. 
\label{part}
\end{equation}
Choosing ${X}^{\nu}=\delta_{t}^{\nu}$
and integrating Eqn.(\ref{part}) in a 3-volume $V$,
\begin{equation}
\int_{V}\partial_t(\sqrt{-\tilde{g}}\widetilde{T}_{t}^{t})%
d^{3}x  
+
\int_{V}\partial_i(\sqrt{-\tilde{g}}\widetilde{T}_{t}^{i})%
d^{3}x
=0
\label{k1}.
\end{equation}
Defining the energy of the perturbations by
$\widetilde{E}=\int_{V}\sqrt{-\widetilde{g}}$ $\widetilde{T}_{t}^{t}d^{3}x$,
and using Gauss theorem,
Eqn.(\ref{k1}) can be written as
\begin{equation}
\frac{d\widetilde{E}}{dt}=-\int_{\Sigma} d\Sigma_i  \sqrt{-\widetilde{g}}\text{
}\widetilde{T}_{t}^{i} .
\label{integral}
\end{equation}
This is the key result obtained by Moncrief in \citep{Moncrief1980}. By a proper choice of the surface $\Sigma$ enclosing the volume $V$, and using the properties of the fields on $\Sigma$, it is possible to determine the sign of the integral without solving it, thus determining if the system is linearly stable. We shall use this method to determine if the model presented in the next section is stable.
 
\section{Accretion of a nonlinear scalar field onto a Schwarzschild black hole}
\label{fro} 
\begin{figure}[h]
       \centering  
       \includegraphics[width=0.4\textwidth]{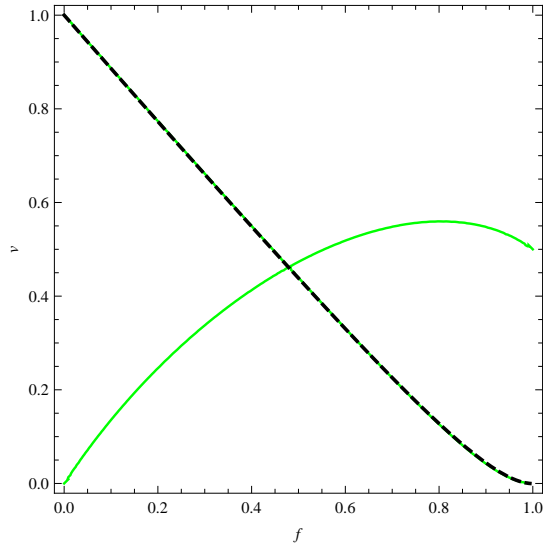}
       \caption{The figure shows in green two trajectories (for $A=3/4$) that pass through the sonic horizon $r_{\rm s}$ (located at the intersection of the curves). They correspond to $\alpha = 1.51833455$. The physically relevant (and unique) trajectory for the problem at hand is the one that starts at infinity ($f=1$) with zero radial velocity and reaches Schwarzschild's horizon ($f=0$) passing through the sonic horizon. Also shown is the plot of the analytic form for the relevant trajectory, to be used in Sect.\ref{efpot} (dotted black curve).}
       \label{ajuste}
\end{figure}

The model studied in \citep{Frolov2004} (called ``ghost condensation") was proposed in 
\citep{ArkaniHamed2003}, and 
consists of a nongravitating nonlinear scalar field with Lagrangian
\begin{equation}%
\mathcal{L}%
(W)=\frac{1}{2}(W-A)^{2},
\label{lag}
\end{equation}
(where $A$ is a constant to be determined later)
in the background of a Schwarzschild black hole, 
\begin{equation}
ds^{2}=fdt^{2}-f^{-1}dr^{2}-r^{2}d\Omega^{2},
\label{schwmet}
\end{equation}
with
\begin{equation}
f(r)=\left(  1-\frac{2M}{r}\right).
\end{equation}
From the assumption of a stationary background and spherical symmetry, it follows that
\begin{equation}
\phi_{0}(t,r)=t+\psi_0\left(  r\right).
\end{equation}
Using Eqn.(\ref{schwmet}) in 
\ref{eom}, the EOM for $\psi_0$ is 
$$
\partial^*_r(r^2{\cal L}_W\partial^*_r\psi_0)=0,
$$
where 
$\partial^{*}_r \equiv f(r)\partial_r$. This equation 
can be integrated to yield
\begin{equation}%
\mathcal{L}%
_{W}\text{ }\partial^{*}_{r}\psi_0=\alpha\frac{r_{g}^{2}}{r^{2}},
\end{equation}
where $\alpha$ is an integration constant, which 
is related to the accretion rate through $\dot M = 
2\alpha \times 4\pi r_{\rm g}^2M^4$ \citep{Frolov2004}. \\
Using the variables  
$v\equiv \partial^{*}_{r}\psi_0$ and $f$, and the Lagrangian given in Eqn.(\ref{lag}), this equation can be written as
\begin{equation}
\left(\frac{1-v^2}{f}-A\right)v = \alpha (1-f^2).
\label{vf}
\end{equation}
As shown in \citep{Frolov2004}, there is only one trajectory that goes from an homogeneous solution at infinity, passes through a sonic horizon (which is the 2-surface of constant $r$ such that 
$v_\perp^2 = c_s^2$, where $c_s$ is the velocity of sound \citep{Bilic1999}), and reaches the Schwarzschild horizon
$r_{\rm g}$. This trajectory is depicted in 
Fig. \ref{ajuste}. It follows from the figure and the definitions of $v
$ and $f$ that $\partial_r\psi_0 \geq 0$, a piece of information that will be used below. 

\section{Stability using the time variation of the energy of the perturbations}

To evaluate the stability of the system reviewed in the last section using the integral 
given in Eqn.(\ref{integral}), it is convenient to choose 
the volume $V$ as that encompassed between the surfaces $r=r_{\rm s}$ and $r \rightarrow \infty$. Since the fields are such that \begin{equation}
\sqrt{-\widetilde{g}}\;\widetilde{g}^{rr}\rightarrow\sqrt{-g}g^{rr}=r^{2}%
\sin\theta,\;\;\;\;\;\widetilde{g}^{rt}\rightarrow 0\;\;{\mbox{for}}\;\;r\rightarrow \infty,
\end{equation}
the integral at infinity reduces to
\begin{equation}
I_{1}=-\int_{\Sigma_\infty}\left.  \sqrt{-\tilde{g}}\;\phi_{1,r}\phi_{1,t}\widetilde{g}%
^{rr}\right\vert _{r\rightarrow\infty}d\Sigma_{r}.
\end{equation}
The assumption of the finiteness of the energy of the perturbations, given by 
\[
\widetilde{E}=\int_{V}\sqrt{-\widetilde{g}}\;\widetilde{T}_{t}^{t}d^{3}x,
\]
leads to the following behaviour of the perturbations at infinity:
\begin{equation}
\phi_{1,t}\sim\frac{a}{r^{\frac{3}{2}+\epsilon}};\;\;\;\;\;\phi_{1,r}\sim\frac
{a^{\prime}}{r^{\frac{3}{2}+\epsilon}},\nonumber
\end{equation}
where $\epsilon >0$, and $a$ and $a'$ are constants.
It follows that $I_1$ is zero. The integral at the sonic horizon is given by
\begin{equation}
I_{2}=%
{\displaystyle\int\limits_{\Sigma_{r_{\rm s}}}}
\left.  \sqrt{-\tilde{g}}\left[  \phi_{1,r}\phi_{1,t}\widetilde{g}^{rr}%
+(\phi_{1,t})^{2}\widetilde{g}^{rt}\right]  \right\vert _{r=r_{\rm s}}d\Sigma_{r}.
\end{equation}
Taking into account that 
$\widetilde{g}^{rr}(r_{\rm s})=0$, we get for the time derivative of the energy of the perturbations
\begin{equation}
\frac{d\widetilde{E}}{dt}=I_{2}=\int_{\Sigma_{r_{\rm s}}}\left.  \sqrt{-\tilde{g}}(\phi_{1,t}%
)^{2}\widetilde{g}^{rt}\right\vert _{r_{\rm s}}dS_{r}.
\end{equation}
From the definition of the effective metric, Eqn.(\ref{defeffmet}), 
it follows that
$
\sqrt{-\widetilde{g}}\;\widetilde{g}^{tr} = - 2 \sqrt{-g}\;\psi_{0,r}.
$
Hence,
$$
\frac{d\widetilde{E}}{dt}=-2\left.\int_{\Sigma_{r_{\rm s}}}r^2\sin\theta \psi_{0,r}\phi_{1,t}^2
d\theta d\varphi\right|_{r=r_{\rm s}}.
$$
Since, as mentioned above, Fig.\ref{ajuste} shows that 
$\psi_{0,r}$ is positive, we conclude that the system is stable.

\section{Stability using the effective potential}
\label{efpot}
\begin{figure}[h]
       \centering  
       \includegraphics[width=0.4\textwidth]{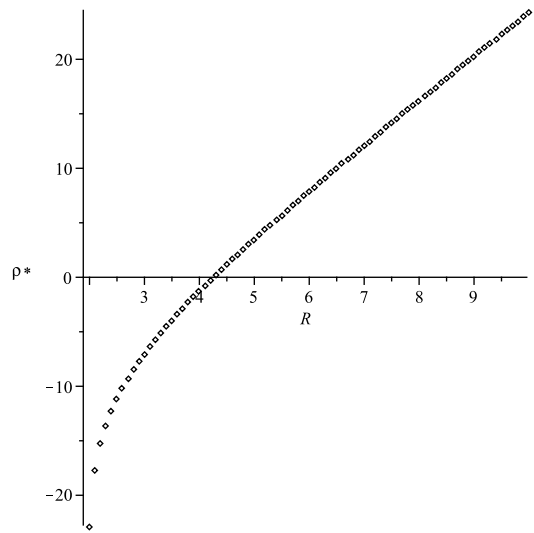}
       \caption{The coordinate $\rho^*$ takes the sonic horizon to $-\infty$, while preserving $+\infty$. In the figure, $R=r'/2M$.}
       \label{rho}
\end{figure}

In this section we shall pursue the more traditional test of the stability of a system, that involves the effective potential (see for instance \citep{Wald1979}). First, a 
change of coordinates is carried out to diagonalize the metric:
\begin{eqnarray}
dt  & = & dt'-\frac{\widetilde{g}_{rt}}{\widetilde{g}_{tt}}dr',\nonumber\\
dr  & = & dr'.
\end{eqnarray}
Only the $tt$ component of the metric is transformed, in such a way that 
the new component is given by 
$$
\widetilde{G}^{t't'}   =  \frac{1}{\widetilde{g}^{rr}}\left(\widetilde{g}^{tt}\widetilde{g}^{rr}%
-\widetilde{g}^{rt}\right).
$$
Starting from the equation of motion 
for the perturbations 
written in the new coordinates, 
\begin{equation}
\partial_{\mu}\left(\sqrt{\widetilde{G}}\widetilde{G}^{\mu\nu}\partial_{\nu}%
\phi_{1}\right)=0,
\end{equation}
and using the decomposition
\begin{equation}
\phi_{1}(t,r,\theta , \phi ) =e^{-i\omega t'}\beta(r')Y_{lm}(\theta,\varphi),
\end{equation}
we get after a long and straightforward calculation,
\begin{equation}
\frac{d^{2}\beta}{d\rho_{\ast}^{2}}+\left[\omega^{2}-V_{\rm eff}(r')\right]\beta=0,
\label{eqradial}
\end{equation}
where we have introduced the coordinate $\rho^*$, defined by 
\begin{equation}
d\rho^{\ast}=F
\mathcal{L}_{W}dr'.
\end{equation}
As shown in Fig.\ref{rho}, $\rho^*$ is 
is very much like the 
tortoise coordinate used in the case of Schwarzschild's black hole.
Notice that in order to build this plot
(as well as that of the effective potential, see below),
an explicit form of the background solution was needed. It was represented by the function
$$
v = \frac{(f-1)^2}{1-a_1f}-a_2f,
$$
where $a_1=0.8599$, and $a_2=0.0003$.
This curve is in very good agreement with that 
obtained from Eqn.(\ref{vf}), as the plots in Fig. 
\ref{ajuste} show.

\begin{figure}[h]
       \centering  
       \includegraphics[width=0.5\textwidth]{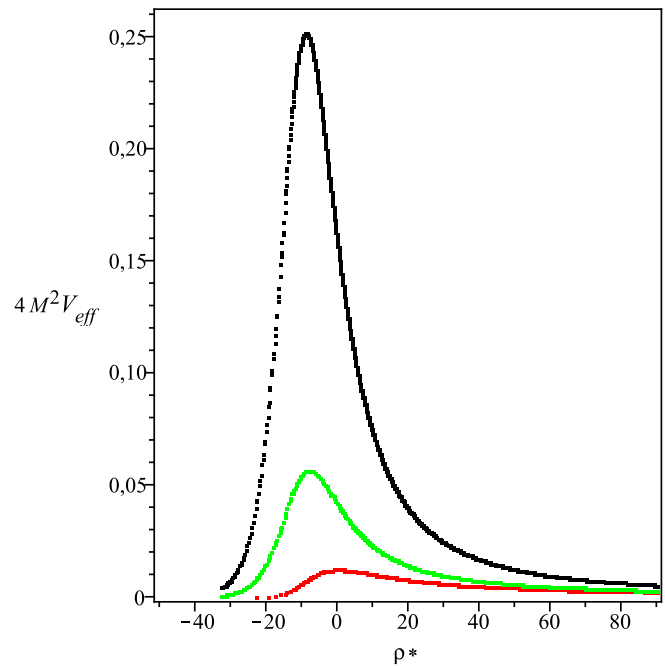}
       \caption{The effective potential as a function of 
$\rho^*$ for $\ell = 0$ (bottom curve), $\ell = 2$ (middle curve), and $\ell = 5$.}
       \label{poteff}
\end{figure}

The effective potential $V_{\rm eff}$ in Eqn.(\ref{eqradial})
is 
given by 
\begin{equation}
V_{\rm eff} = \frac{\ell (\ell +1)\sigma}{r'^2 G^{t't'}}
-\frac 1 4 \left[\frac{1}{r'^2}\frac{d}{d\rho^*}\left(\frac{r'^2}{\sigma}\right)
\right]^2-
\frac 1 2 
\frac{d}{d\rho^*}\left[\frac{1}{r'^2}\frac{d}{d\rho^*}\left(\frac{r'^2}{\sigma}
\right)\right],
\end{equation}
where $\sigma \equiv \sqrt{M^{rt}-M^{tt}M^{rr}}$.
This expression reduces to the usual one in the case of a linear theory for the scalar field, see for instance \citep{Frolov1998}.
Fig. \ref{poteff} displays the plot of $V_{\rm eff}(\rho^*)$ for several values of the angular momentum $\ell$.
As discussed for instance in \citep{Wald1979}, the positivity of the potential is sufficient to guarantee the stability of the sistem.

\section{Conclusions}
\label{disc}
We have shown that the system composed of a Schwarzschild black hole 
accreting a steady-state and spherically symmetric ghost condensate described by the Lagrangian 
given in Eqn.(\ref{lag}) is linearly stable using two methods. The method developed by Moncrief consists in determining the sign 
of the time derivative of the energy of the perturbations through a surface 
integral, and led to the linear stability of the system in a few steps. 
This method profits from the symmetries of the system, 
and it does not use the explicit form of the solution for the scalar field.  
We have also shown that the system is linearly stable by the method of the effective potential but in this case, much longer calculations are involved, and an explicit form of the solution is required.
Extensions of this work, currently under way, are the study of the
nonlinear stability of the system (since linear stability is only a pre-requisite for full stability), and the generalization of Moncrief's method to other Killing vectors.

\begin{acknowledgments}
SEPB would like to acknowledge support from FAPERJ, CNPQ and UERJ,

\end{acknowledgments}

\bibliographystyle{plain} 
\bibliography{bibliography} 

\end{document}